\documentclass[conference]{IEEEtran}
\IEEEoverridecommandlockouts

\usepackage{amsthm}
\usepackage{inputenc}
\newtheorem{theorem}{Theorem}
\newtheorem*{theorem-non}{Theorem}

\usepackage{graphicx}

\usepackage{algpseudocode}
\usepackage{algorithm}
\usepackage[algo2e]{algorithm2e} 
\usepackage{amsmath,amssymb,amsfonts}

\usepackage{booktabs}       
\usepackage{multirow}

\usepackage{caption}

\def\BibTeX{{\rm B\kern-.05em{\sc i\kern-.025em b}\kern-.08em
    T\kern-.1667em\lower.7ex\hbox{E}\kern-.125emX}}
\begin{document}

\title{Supervised clustering of high dimensional  data using regularized mixture modeling}

\author{
\IEEEauthorblockN{Wennan Chang}
\IEEEauthorblockA{Department of Electrical and\\
Computer Engineering\\
Purdue University\\
}\\   
\IEEEauthorblockN{Chi Zhang*}
\IEEEauthorblockA{Medical and Molecular Genetics\\
Indiana University School of Medicine\\
Email: czhang87@iu.edu}
\and
\IEEEauthorblockN{Changlin Wan}
\IEEEauthorblockA{Department of Electrical and\\
Computer Engineering\\
Purdue University\\
}\\ 
\IEEEauthorblockN{Sha Cao*}
\IEEEauthorblockA{Department of Biostatistics\\
Indiana University\\
Email: shacao@iu.edu}
\and
\IEEEauthorblockN{Yong Zang}
\IEEEauthorblockA{Department of Biostatistics\\
Indiana University\\
}\\                 
}

\maketitle

\begin{abstract} Identifying relationships between molecular variations and their clinical presentations has been challenged by the heterogeneous causes of a disease. It is imperative to unveil the relationship between the high dimensional molecular manifestations and the clinical presentations, while taking into account the possible heterogeneity of the study subjects.  We proposed a novel supervised clustering algorithm using penalized mixture regression model, called CSMR, to deal with the challenges in studying the heterogeneous relationships between high dimensional molecular features to a phenotype. The algorithm was adapted from the classification expectation maximization algorithm, which offers a novel supervised solution to the clustering problem, with substantial improvement on both the computational efficiency and biological interpretability. Experimental evaluation on simulated benchmark datasets demonstrated that the CSMR can accurately identify the subspaces on which subset of features are explanatory to the response variables, and it outperformed the baseline methods. Application of CSMR on a drug sensitivity dataset again demonstrated the superior performance of CSMR over the others, where CSMR is powerful in recapitulating the distinct subgroups hidden in the pool of cell lines with regards to their coping mechanisms to different drugs. CSMR represents a big data analysis tool with the potential to resolve the complexity of translating the clinical manifestations of the disease to the real causes underpinning it. We believe that it will bring new understanding to the molecular basis of a disease, and could be of special relevance in the growing field of personalized medicine.
\end{abstract}

\section{Introduction}

Detection and estimation of the molecular markers associated with phenotypic features is one of the most important problems in biomedical research. Predicative models have been extensively used to link molecular markers to a phenotypic trait, however, the unobserved patient heterogeneity obfuscates the effort to build a unified model that works for all hidden disease subtypes. It has been well understood that various subtypes exist for many common diseases, which vary in etiology, pathogenesis, and prognosis \cite{curtis2012genomic, schlicker2012subtypes, guinney2015consensus}. For example, the cancer cells are constantly evolving in the tumor microenvironment, and they may acquire variations on alternative pathways in response to treatment, which explains why certain patients have better prognoses than others in response to the same treatment \cite{marusyk2010tumor}. This implies that the same predicative model that links molecular markers to a phenotypic trait may not be valid for every patient, and further it is unclear to what extent the patients should be considered together \cite{kobel2008ovarian}. Therefore, it is judicious to construct a set of heterogeneous models, each of which corresponds to one subtype.

\begin{figure*} 
    \centering
    \includegraphics[width=17cm]{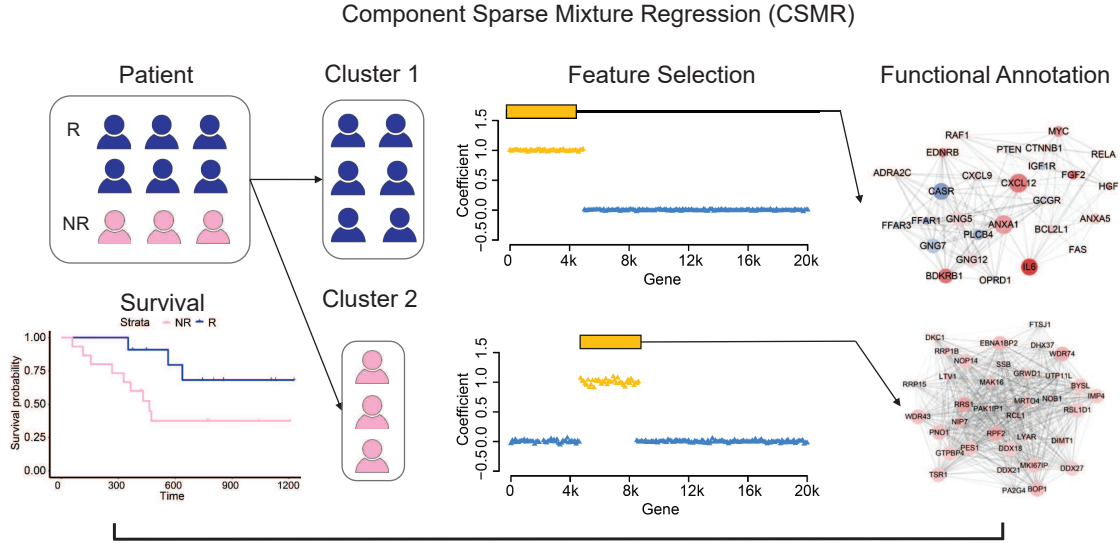}
    \caption{The motivation of CSMR. Under the same treatment, some patients acquired one mechanism to deal with the drug, (blue), while others picked up another (pink), resulting in different prognoses to the same treatment. The motivation of CSMR is to cluster the patients in a supervised fashion and examine what are the genes (yellow) that are selected in tumor progression that lead to the different drug resistance subtypes of patients, and their functions (network).}
\end{figure*}

The fast advancement in high-throughput technology has transformed the biomedical research ecosystem by scaling data acquisition, providing us with unprecedented opportunity to interrogate biology in novel and creative ways. For cancer research, the Broad Institute Cancer Cell Line Encyclopedia (CCLE) \cite{barretina2012cancer}, Cancer Therapy Response Portal (CTRP) v1/v2 \cite{basu2013interactive}, and Genomics of Drug Sensitivity in Cancer (GDSC) \cite{yang2012genomics} datasets contain 24, 185, 481, and 261 drug compound screening data for 504, 242, 860, and 1001 cell lines respectively, together with the multi-omic profiles of the cell lines; the Cancer Genome Atlas (TCGA) has collected biospecimens and matched clinical phenotypes for over 10,000 cancer patients \cite{weinstein2013cancer}. Consequently, for each sample, there is a tremendous amount and variety of data: ~20,000 genes’ expression profiles, 1 million single-nucleotide polymorphism genotypes, exome and whole-genome sequences, methylation of tens of thousands of CpG islands and the expression of microRNA. From this plurality of data, we anticipate that exploratory methods will serve to extract and characterize molecular subgroups relevant to phenotypic outcomes. However, the growing number of variables does not necessarily increase the discriminative power in classification \cite{fan2014challenges}. To identify the most important molecular biomarkers, variable selection is one of the most commonly used approaches. In particular, the penalized regularization methods have received a great deal of attention \cite{tibshirani1996regression, zou2006adaptive, fan2001variable}. Despite major progress in the research of penalized regression, heterogeneity in variable selection of high dimensional feature spaces remains to be challenging. 

Unsupervised learning algorithms such as finite mixture models are typically employed to deal with heterogeneity in subpopulations by assuming a separate distribution for each subpopulation \cite{mclachlan2004finite, fraley1998many, fraley2002model}. Based on the molecular subtypes, deeper investigation into the molecular and phenotypic distinctions within each subtype could be carried out. Although the clustering methods may produce satisfactory classification of subtypes, it does not select molecular markers distinctive for each subtype, which however is essential in precision medicine. And more importantly, without any supervision, the defined clusters based on a sea of molecular features may not necessarily relate to the phenotype of interest. 

Hence, to study the heterogeneous relations of molecular markers to a certain phenotypic trait, our challenges is distinct in two ways: the variables of interest to each subgroup may be a distinct and sparse set of the high dimensional molecular features, and the set of patients in each subgroup is not known. In this article, we proposed a novel and efficient supervised clustering algorithm based on penalized mixture regression model that synergizes with potential heterogeneity in high dimensional regression problem. Essentially, we assume that observations belong to unlabeled classes with class-specific regression models relating their unique and selective molecular markers to the phenotypic outcome.

\section{Preliminaries}
Since first introduced in \cite{goldfeld1973estimation}, finite mixture Gaussian regression (FMGR) has been extensively studied and widely used in various fields \cite{bohning1999computer, hennig2000identifiablity, jiang1999hierarchical, mclachlan2004finite, xu1996convergence, fruhwirth2006finite}. Let $Y=(y_1,...,y_N)^T \in \mathcal{R}^N$, $X= [ \boldsymbol{x}_1,...,\boldsymbol{x}_N ]^T \in \mathcal{R}^{N \times (P+1)}$ be a finite set of observations, and $X$ the design matrix with intercept and $P$ independent variables, and $Y$ the response vector. Consider an FMGR model parameterized by $\boldsymbol\theta = \{ (\pi_k, \boldsymbol{\beta}_k, \sigma_k^2) \}_{k=1}^K$, it  is assumed that when the $i$-th observation, $(\boldsymbol{x}_i, y_i)$, belongs to the \textit{k}-th component, $k=1,...,K$, then $y_i = \boldsymbol{x}_i^T \boldsymbol{\beta}_k +\epsilon_{ik}$, and $\epsilon_{ik}  \sim N(0, \sigma_k^2)$. In other words, the conditional density of $y$ given $\boldsymbol{x}$ is $f(y| \boldsymbol{x,\theta}) = \sum_{k=1}^K \pi_k \mathcal{N}(y; \boldsymbol{x^T \beta_k}, \sigma_k^2)$, where $\mathcal{N}(y; \mu, \sigma^2 )$ is the normal density function with mean $\mu$ and variance $\sigma^2$. And the log-likelihood for observations $\{(\boldsymbol{x}_i, y_i)\}_{i=1}^N$ is 
\begin{equation}
\mathcal{L}(\boldsymbol\theta)=\Sigma_{i=1}^N \log(\Sigma_{k=1}^K \pi_k \mathcal{N}(y_i; \boldsymbol{x}_i^T \boldsymbol{\beta}_k, \sigma_k^2 ))
\end{equation}

The unknown parameters $\boldsymbol\theta$ can be estimated by the maximum likelihood estimator
(MLE), which maximizes (1). Note that the maximizer of (1) does not have an explicit solution and is usually solved by the EM algorithm. Basically, the EM algorithm maximizes the complete log likelihood function, $\mathcal{L}^c(\boldsymbol\theta)$, through iterative steps, which is defined by
\begin{equation}
\mathcal{L}^c(\boldsymbol\theta)=\Sigma_{i=1}^N \Sigma_{k=1}^K z_{ik}[ \log \pi_k+ \log\mathcal{N}(y_i; \boldsymbol{x}_i^T \boldsymbol{\beta}_k, \sigma_k^2 )]
\end{equation}
where $z_{ik}$ is a cluster indicator variable, and $z_{ik}=1$ if the $i$-th observation belongs to the $k$-th cluster, and 0 otherwise. 

While mixture regression model is capable of handling the heterogeneous relationships, it doesn't work in the case of high dimensional molecular features, where the total number of parameters to be estimated is far more than the total number of observations. In addition, with the dense linear coefficients given by the ordinary EM algorithm, it is hard to deduce the disease subtype specific molecular markers and make meaningful interpretations. 

Penalized mixture regression has been explored in different settings \cite{ khalili2007variable,stadler2010L,fan2010comments,lloyd2018globally,li2019drug} to handle the high dimensional mixture regression problem. The variable selection problem in finite mixture of regression model was first studied using regularization methods such as LASSO \cite{tibshirani1996regression} and SCAD \cite{fan2001variable} in \cite{khalili2007variable}, called FMRS. They considered the traditional cases when the number of candidate covariates is much smaller than the sample size, and proposed a modified expectation–maximization (EM) algorithm to perform both estimation and variable selection simultaneously.  In \cite{stadler2010L}, the authors proposed a reparameterized mixture of regressions model, and showed evidence for the advantage of multiple components that can be exploited for variable selection over non-mixture linear regression. A block-wise Minorization Maximization (MM) algorithm was proposed in \cite{lloyd2018globally}, where at each iteration, the likelihood function is maximized with respect to a block of variables while the rest of the blocks are held fixed. To solve the population heterogeneity and feature selection problems, an imputation-conditional consistency (ICC) algorithm was proposed by \cite{li2019drug}, resulting in consistent estimators. While some of the methods may produce consistent estimates of $\boldsymbol{\theta}$ under proper conditions, they tend to suffer from slow convergence rate in high dimensional setting, especially with smaller $N$ or larger $K$, and the number of hyper-parameters for regularization further drags down the computational efficiency caused by the need of cross validation. We here propose a novel algorithm based on classification EM for penalized mixture regression to circumvent these existing challenges in clustering high dimensional data using mixture regression, which largely increased the computational efficiency. 

The rest of the article is organized as follows: in Section 3, we introduce our algorithm, \textbf{C}omponent-wise \textbf{S}parse \textbf{M}ixture \textbf{R}egression (\textbf{CSMR}); in section 4, we compare CSMR with four state-of-the-art algorithms on synthetic datasets, 
in section 5, we applied all the five algorithms on 24 drug sensitivity data in CCLE, to screen for genes that underlie the heterogeneous drug resistance mechanisms.

\section{Methods}
We assume that the samples belong to different sub-populations, each of which is defined by a distinct relationship between the molecular biomarkers to the phenotype of interest, and the molecular markers are sparse subsets of the high dimensional molecular profiles specific to each sub-population. Figure 1 illustrated an example where the patients fall under two distinct subgroups: blue for patients acquiring one mechanism to the treatment that resulted in responsiveness, while pink for patients acquiring another mechanism to the same drug that resulted in non-responsiveness. The goal of our method is to cluster the samples (blue and pink) supervised by the patients drug sensitivity measure, and find the defining molecular features (yellow) associated with each cluster. The identified molecular features could be further studied to guide targeted therapeutic designs. 

\subsection{The penalized likelihood of mixture regression}

Knowing that $\boldsymbol{\beta}_k$ is sparse means many elements in $\boldsymbol{\beta}_k$ will tend to be close to zero, but not exactly zero without proper regularization in the model. To simultaneously shrink the insignificant regression coefficients in  $\boldsymbol{\beta}_k$ and estimate $\boldsymbol\theta$, we introduce penalty term to (1) and optimize the following penalized log likelihood function;
\begin{equation}
\max_{\boldsymbol\theta\in \boldsymbol\Theta_0}{\mathcal{L}(\boldsymbol\theta)- P_{\boldsymbol\lambda}(\boldsymbol\theta)}
\end{equation}
where $\mathcal{L}(\boldsymbol\theta)$ denotes the observed log likelihood, and $P_{\boldsymbol\lambda}(\boldsymbol\theta): \mathcal{R}^P \rightarrow \mathcal{R}$ is a regularizer of the regression coefficients, and the penalty for each component is dependent on a component specific hyperparameter $\lambda_k>0$. For notational convenience, we define $\mathcal{L}^p(\boldsymbol\theta)={\mathcal{L}(\boldsymbol\theta)- P_{\boldsymbol\lambda}(\boldsymbol\theta)}$. Various types of penalty were used in mixture regression model \cite{li2019drug,khalili2007variable}, but we consider LASSO penalty form as it is convex and thus advantageous for numerical computation \cite{tibshirani1996regression}, i.e., 
\begin{equation}
P_{\boldsymbol\lambda}(\boldsymbol\theta)=\Sigma_{k=1}^K \pi_k \Sigma_{j=1}^P \lambda_k |\beta_{jk}| 
\label{eq:1}
\end{equation}

Similar to the case of low dimensional mixture regression, EM algorithm could be adopted by maximizing the penalized complete log likelihood function $$\mathcal{L}^{pc}(\boldsymbol\theta)={\mathcal{L}^c(\boldsymbol\theta)- P_{\boldsymbol\lambda}(\boldsymbol\theta)}$$
by iterating between the following E-step and M-step:

\noindent E-step: computing the conditional expectation of $\mathcal{L}^{pc}(\boldsymbol\theta)$ with respect to $z_{ik}$ given the current estimates $\boldsymbol\theta^{(m)}$. The conditional expectation is
\begin{equation}
Q(\boldsymbol\theta; \boldsymbol\theta^{(m)})=\Sigma_{i=1}^N \Sigma_{k=1}^K p^{(m)}_{ik}[ \log \pi_k+ \log\mathcal{N}(y_i; \boldsymbol{x}_i^T \boldsymbol{\beta}_k, \sigma_k^2 )]-P_{\boldsymbol\lambda}(\boldsymbol\theta)
\end{equation}

Then the conditional expectation of $z_{ik}$ is given by 
\begin{equation*}
p_{ik}^{(m)} = 
\frac{\pi_k^{(m)} \mathcal{N} (y_i; {\boldsymbol{\beta}^{(m)}_k}^T \boldsymbol{x}_i, \sigma_k^{2^{(m)}}) }
{\sum_{l=1}^K \pi_l^{(m)} \mathcal{N} (y_i; {\boldsymbol{\beta}^{(m)}_l}^T \boldsymbol{x}_i, \sigma_l^{2^{(m)}}) }
\end{equation*}

\noindent M-step: maximizing $Q(\boldsymbol\theta; \boldsymbol\theta^{(m)})$ with respect to $\boldsymbol\theta$, i.e., 
\begin{equation}
\boldsymbol\theta^{(m+1)}=\{\pi_k^{{m+1}}, \boldsymbol{\beta}_k^{(m+1)}, \sigma_k^{2^{(m+1)}}\}_{k=1}^K=\max_{\boldsymbol\theta\in \boldsymbol\Theta_0}{Q(\boldsymbol\theta; \boldsymbol\theta^{(m)})}
\end{equation}
Unlike the low-dimensional case, where $\pi_k^{{m+1}}, \boldsymbol{\beta}_k^{(m+1)}, \sigma_k^{2^{(m+1)}}$ all have closed form solutions, maximizing (6) is more complicated due to the involvement of $\pi_k, \boldsymbol{\beta}_k$ in the penalty term and the non-differentiable form of $P_{\boldsymbol\lambda}(\boldsymbol\theta)$ at $\beta_{jk}=0$.

\subsection{The classification EM algorithm}
The Classification Expection Maximization (CEM) algrorithm is an variant of the EM algorithm. It has been popularly used in the Finite Gaussian Mixture Model\cite{blomer2016hard,celeux1992classification}, and shown to be have faster convergence rate \cite{faria2010fitting}. Basically, the assignments $\{ z_i \}_{i=1}^N$ define a partition $\mathcal{C} = \bigcup_{k=1}^K C_k$ s.t. $i \in C_k$ iff $z_i = k$. The CEM algorithms maximizes $\mathcal{L}^{c}(\boldsymbol\theta)$ through iterating among three steps: \\

\noindent E-Step: calculating conditional expectation of $p_{ik}^{(m)}$, similar to the traditional EM.

\noindent C-Step: disentangle the observations into $K$ classes, by assigning $C_k^{(m+1)}$ as the set of observations most likely in cluster $k$, i.e., $\{ i|k = \underset{l \in \{ 1,...,K \}}{\operatorname{argmax}} \ p_{il}^{(m)}, i=1,...,N\}$. Let $n_k$ denotes the total number of observations in cluster $k$.

\noindent M-Step: parameter estimation within each disentangled cluster, where $\pi_k^{(m+1)}$ is estimated as $n_k^{(m+1)}/N$, and $\boldsymbol{\beta}_k^{(m+1)}, \sigma_k^{2^{(m+1)}}$ are simply estimated as the ordinary least square (OLS) estimators using observations in $C_k^{(m+1)}$ only.

We show the convergence of the CEM algorithm for the low-dimensional case in Theorem 1.
\begin{theorem} For the sequence $\mathcal{C}^{(m)}, \boldsymbol{\theta}^{(m)}$ updated as CEM, the complete data likelihood converges to a stationary value. Moreover, if the maximum likelihood estimates of the parameters are well-defined, the sequence $\mathcal{C}^{(m)}, \boldsymbol{\theta}^{(m)}$ also converges to a stationary position.
\end{theorem}

The biggest advantage of the CEM algorithm is that it disentangles the mixture into individual non-overlapping components, such that flexible sparsity control could be easily achievable within each component. Hence for the high dimensional mixture regression problem, we could simply replace the OLS estimator in the M step of the CEM algorithm by a sparse estimator, i.e., 

$$
\underset{\boldsymbol{\beta}_k, \sigma_k^2}{\textbf{argmax}}
\underset{i \in \mathcal{C}_k^{(m+1)}}{\sum}
\log\mathcal{N}(y_i; \boldsymbol{x}_i^T \boldsymbol{\beta}_k, \sigma_k^2 ) - \lambda_k \pi_k \Sigma_{j=1}^P |\beta_{jk}| 
$$ 
This is simply $L_1$ regularized linear regression, for which many efficient algorithms exist \cite{hastie2014glmnet}.

\subsection{The CSMR algorithm}
Here we proposed the CSMR algorithm to solve the high dimensional mixture regression problem based on the CEM algorithm. In CSMR, the mixture regression setting could handle the hidden cluster problem, and the disentangled clusters under CEM could efficiently solve the feature selection problem in high dimensional setting. At E-step, we calculate the posterior probability $p_{ik}$ similar to traditional EM and ECM; at C-step, we assign each observation to a cluster that it most likely belongs to, similar to traditional CEM; at the M-step, for each component, we perform regularized linear regression to obtain a sparse set of non-zero coefficients.

A big challenge with the penalized mixture regression problem is the choice of component specific penalty parameters $\lambda_k$. The $\lambda_k$'s are related to the amount of regularization, and their selection is a critical issue in a penalized likelihood approach. It is usually based on a trade-off between bias and variance: large values of tuning parameters tend to select a simple model whose parameters estimates have smaller variance, whereas small values of the tuning parameters lead to complex models, with smaller bias. Cross-validation over a grid search is the commonly adopted method to select the optimal combination of $\lambda_k$, but this becomes increasingly prohibitive with the increase of $K$, especially when we don't have a good knowledge of the theoretical range of the $\lambda_k$. 

Hence, instead of first performing penalized linear regression for given $\lambda_k$ and then search the optimal combination of $\lambda_k$ \cite{khalili2007variable}, we propose to conduct the tuning of $\lambda_k$ with cross validation inside the ECM iterations. Specifically, under the CEM algorithm, all the components are disentangled, we could hence perform hyperparameter tuning inside each iteration within each component. This is to say, at the M-step, we not only estimate the regression coefficients, but also find the best tuning parameter $\lambda_k$ for the component. Hence, at the end of the algorithm, we avoid the hyperparameter tuning, as they have already been selected within the iteration. We adopted efficient cross validation algorithm for selecting the optimal regularization parameter under $L_1$ regularized linear regression \cite{hastie2014glmnet}. Since we no longer need to run the algorithm multiple times on a $K$-dimensional grid space of the penalty parameters, and could hence largely reduce the computational cost. We have shown in simulation studies that penalty parameters selected this way empirically worked very well.

Another adaptation on the traditional ECM algorithm of CSMR is a model refit step following the ECM steps. To increase the numerical stability and achieve faster convergence, at the end of each iteration, we refit the mixture regression model using flexible EM algorithm with only the selected variables of each component. Basically, for each component, the coefficients of the variables not selected at the M-step will be forced to be zero. This could be easily achievable by allowing only the selected variables of component $k$ to enter into the model fitting of the $k$-th regression parameters.

\begin{algorithm}
\SetAlgoLined
\textbf{Input:} $X_{N \times P}, Y_{N \times 1}, K$ \par
\textbf{Output:} $\boldsymbol{\theta}, \mathcal{C} = \bigcup_{k=1}^K \mathcal{C}_k, \{ \beta_{0k}, \boldsymbol{\beta}_k \}_{k=1}^K$ \par
\textbf{Initialization:} $\boldsymbol{\theta}^{(0)}=\{\pi_k^{{0}}, \boldsymbol{\beta}_k^{(0)}, \sigma_k^{2^{(0)}}\}_{k=1}^K$ \par
 \For{m=0,...,Max Iteration}{
    E-step: Compute the conditional expectation of $z_{ik}$ similar to traditional EM algorithm.\\
    C-step: For $k=1,...,K$, assign $C_k^{(m+1)}$ as the set of observations that are mostly likely in component $k$. \\
    M-step: For $k=1,...,K$, the relative cluster size is updated by $\hat\pi_k^{(m+1)} = \frac{n_k^{(m+1)}}{N}$, and the tuning parameter $\lambda_k^{(m+1)}$, and regression parameters $( \hat{\boldsymbol{\beta}}_k^{(m+1)}, \hat\sigma_k^{(m+1)} )$ are selected and estimated using cross validation, such as the cv.glmnet function in glmnet package. 

    Model refit: refit the FMGR model by allowing only the selected variables in each component and to obtain $\{\pi_k^{(m+1)}, \boldsymbol{\beta}_k^{(m+1)}, \sigma_k^{(m+1)}\}_{k=1}^K$ given by this flexible modeling\\
    Stop if converged.
 }
 \caption{CSMR}
\end{algorithm}

The CSMR algorithm requires the initialized values $\boldsymbol{\theta}$. Here, we order the features based on its individual Pearson correlation with the response variable, and then fit a low-dimensional mixture regression model solved by traditional EM algorithm using the top correlated genes. CSMR is implemented in R, and was made available in https://github.com/zcslab/CSMR.

\subsection{Selection of component number $K$}
The number of clusters $K$ is a sensible parameter because it describes the heterogeneity of the population. For selection of $K$, we could use a modified BIC criterior that minimizes
$$
BIC(K)=-2\mathcal{L}^{pc}(\boldsymbol\theta^*_K)+log(N)d_K
$$
where $\boldsymbol\theta^*_K$ represents the parameter estimates for $K$, and $d_K=K+(K-1)+\Sigma_{k=1}^K \Sigma_{j=1}^P 1_{\{\beta_{jk}\neq 0\}}$ is the effective number of parameters to be estimated, similar to  \cite{pan2007penalized}. Specifically, there are $K$ standard deviations, $\sigma_k$, associated with the $K$ regression lines; $K-1$ component proportions, $\pi_k$, since $\Sigma_k \pi_k=1$; and all the non-zero linear regression coefficients for all the $K$ components.

In addition to the BIC criteria, we also offer a cross validation algorithm for the selection of $K$. Take a 5-fold cross validation as an example. For given $K$, at each repetition, 80\% samples are used for training to obtain the regularized parameter $\boldsymbol\theta^*_K$. Then, for a sample $(\boldsymbol{x}_i,y_i)$ drawn from the 20\% testing samples, its cluster membership, $k_0$, is first predicted as

$$k_0=\max_k\pi_{k,K}^*\mathcal{N}(y_i; \boldsymbol{x}_i^T \boldsymbol{\beta}_{k,K}^*, \sigma_{k,K}^{2*} )$$

\noindent Here,  $\pi_{k,K}^*, \boldsymbol{\beta}_{k,K}^*, \sigma_{k,K}^{2*}$ denote the CSMR estimated parameters when the number of components is $K$. After assigning the observation to component $k_0$, we could make prediction of the response based on linear regression, i.e.
$\hat{y_i}=\boldsymbol{x}_i^T \boldsymbol{\beta}_{k_0,K}^*$, as well as the associated residual, $y_i-\hat{y_i}$. Notably, such a prediction of the response is different from simple linear regression, as the prediction process requires knowing the value of the response, in order to assign it to the right cluster. After knowing its cluster membership, a prediction of the response could be made. 

A large $K$ will tend to overfit the data with more complex model of higher variance, while smaller $K$ might select a simpler model with larger bias. Using the independent testing data, we could decide how to balance the trade-off between bias and variance. To evaluate how the estimated model under $K$ explains the testing data, we could calculate the root-mean-square-error between $y_i$ and $\hat{y_i}$, or Pearson correlation between the two. By repeating this procedure for multiple times, a more robust and stable evaluation of the choice of $K$ should be derived based on the summarized RMSE or Pearson correlations.


\section{Application to simulation data}
\subsection{Data generation procedure}
We simulated the independent variables $x_i,i=1,...,P$, which follows i.i.d normal distribution, i.e., $x_{ij} \sim N(0,1)$. The component proportions were made to be equal, i.e., $\pi_k=\frac{1}{K}$. For component $k$, a random sample of size $M_0$ were taken from $\{1,...,P\}$, denoted as $I_k$. And $\beta_{ki}\in Unif((-b,-a)\bigcup(a,b))$, if $\beta_{ki}\in I_k$; $\beta_{ki}=0$, if $\beta_{ki}\not\in I_k$.

The response variable $Y_k$ was generated by the following two-step process:

\noindent 1. Draw component $z_i \in \{ 1,...,K \}$ with probability $ p(z_i=k | \theta) = \pi_k $.

\noindent 2. Draw an observation $y_i$ according to normal distribution $N (\beta_{0k} + \boldsymbol{\beta_k^T x_i}, \sigma_k^2)$.\\

 Here, we fix $a = 2, b = 5,  P=100$. We explored the performances of existing methods under 12 different simulation scenarios, for each of which, 100 repetitions were conducted:\\
Cases 1-3. $N=200,300,400, P = 100, K = 2, \sigma=1, M_0=5$\\
Cases 4-6. $N=400, P = 100, K = 2,3,4, \sigma=1, M_0=5$\\
Cases 7-9. $N=400, P = 100, K = 2, \sigma=0.5,1,2, M_0=5$\\
Cases 10-12. $N=400, P = 100, K = 2, \sigma=1, M_0=5, 8,20$\\

\subsection{Baseline methods}
We compared CSMR with five different methods, including $\mathcal{L}_1$ penalized regression, or LASSO; $\mathcal{L}_2$ penalized regression, or Ridge regression (RIDGE); random forest based regression (RF), FMRS \cite{khalili2011feature} and ICC \cite{li2019drug}. They differ in their ability to perform prediction, clustering and variable selection, as shown in Table 1.

\begin{table}[H]
  \centering
        \caption{Baseline methods}
   \begin{tabular}{|c|c|c|c|}
        \hline
        & Prediction  & Clustering & Variable selection\\
        \hline
     CSMR &$\times$&$\times$&$\times$ \\
        \hline
     LASSO &$\times$&&$\times$ \\
        \hline
     RIDGE &$\times$&& \\
        \hline
     RF &$\times$&&$\times$ \\
        \hline
     FMRS &$\times$&$\times$&$\times$ \\
        \hline
     ICC &$\times$&$\times$&$\times$ \\
        \hline
    \end{tabular}
\end{table}


Among them, CSMR, ICC and FMRS are capable of doing variable selection at the same time of sample clustering. However, FMRS can only deal with relatively lower dimensional features.

\begin{figure} 
    \centering
    \includegraphics[width=8cm]{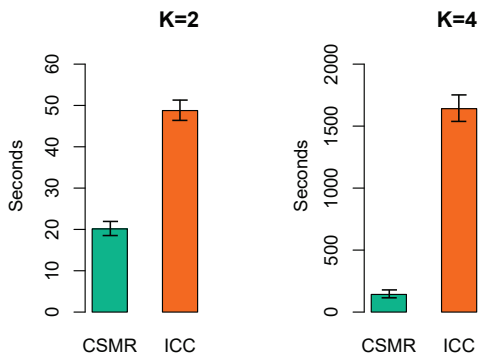}
    \caption{Time consumption of CSMR, and ICC on simulation data for $K=2$ (left) and $K=4$ (right), and $N=400, \sigma=1, M_0=20$ over 100 repetitions, error bars indicate standard deviations.}
\end{figure}

\subsection{Performance comparisons}
We focused on four metrics for method comparisons: 1) the average correlation between predicted and observed response; 2) the true positive rate (TPR) and 3) true negative rate (TNR) of variable selection; 4) the rand index of sample clustering (RI). Note that for observation $i$, its predicted response is given by $ \Sigma_{k=1}^K z_{ik}(\boldsymbol{x}_i^T \boldsymbol{\beta}_k)$, where $z_{ik}$ is its cluster membership indicator. The average of the four metrics over 100 simulations in each scenario was calculated and shown in Table 2. Here, we assume that the true $K$ is known. 

In general, CSMR performs the best in terms of the four evaluation metrics in the majority of the scenarios. For prediction accuracy of the response using correlation, CSMR and ICC perform comparably well, and CSMR slightly better in most of the cases. This is expected as LASSO, RIDGE and RF can not deal with the sample heterogeneity, and FMRS does not work well when the feature dimension is high. For sensitivity and specificity of the variable selection, CSMR performs significantly better than ICC and FMRS. Selection of the right variables is very important as it characterizes the unique features of each component, based on which, we could further deduce the biological interpretation of each unique component.  ICC and FMRS suffer from very low sensitivity of variable selection in almost all cases, and their specificity metrics are not desirable either. For clustering, CSMR again has the best or close to the best performance compared with ICC and FMRS. ICC achieved similar performance with CSMR in some cases, but it clearly suffers when $K$ or the number of effective variables $M_0$ become large. We also compared the computational efficiency of CSMR and ICC under the parameter setting: $N=400, P=100, \sigma=1, M_0=20, K=$2 or 4. Figure 2 shows the computational cost and its standard deviation for two algorithms over 100 repetitions. Clearly, the computational efficiency of ICC drops significantly when $K$ increases from 2 to 4, while the time consumption for CSMR stays approximately the same.

Hence from simulation data, we could see that CSMR achieved the most desirable performance in terms of prediction accuracy, variable selection and clustering, compared with three non-mixture regularized models, and two mixture models. While ICC is competitive in some cases, it severely suffers from poor variable selection, and its computational cost is too prohibitive compared with CSMR. The CSMR has a built-in cross validation step within the CEM iteractions, which could largely increase the sensitivity and specificity of the variable selection procedure, and the flexible model refit step following the CEM steps guarantees that the algorithm could achieve faster convergence and more stable results.

\begin{table*}
  \centering
  \resizebox{1.1\textwidth}{!}{\begin{minipage}{\textwidth}
        \caption{Comparisons of CSMR with other five methods in various simulation settings}
  \begin{tabular}{ccccc|ccc|ccc|ccc}
    \toprule
    \multirow{ 3}{*}{Metrics} & Experiment & \multicolumn{3}{c|}{$\sigma=1,N=400,M_0=5$} & \multicolumn{3}{c|}{$K=2,N=400,M_0=5$} & \multicolumn{3}{c|}{$K=2,\sigma=1,M_0=5$} & \multicolumn{3}{c}{$K=2,\sigma=1,N=400$}\\
                       &   &          & $K$ &  &            & $\sigma$ &  &               & $N$ &    &  & $M_0$ &  \\
    \cmidrule(r){2-14}
                       & Parameter  &         2 & 3 & 4 &           0.5 & 1 & 2 &              200 & 300 & 400   & 5 & 8 & 20\\
    \midrule
                &  \textbf{CSMR} & \textbf{0.992} & \textbf{0.988} & \textbf{0.999} & \textbf{0.998} & \textbf{0.994} & 0.977 & 0.987 & \textbf{0.993} & \textbf{0.994}  & \textbf{0.992} & \textbf{0.995} & \textbf{0.994}  \\
                &  ICC           & 0.992 & 0.985 & 0.909 & 0.998 & 0.984 & \textbf{0.982} & \textbf{0.992} & 0.992 & 0.984    & 0.992 & 0.994 & 0.984 \\
$Cor(y,\hat{y})$&  LASSO         & 0.743 & 0.654 & 0.585 & 0.745 & 0.778 & 0.729 & 0.776 & 0.756 & 0.778  & 0.743 & 0.754 & 0.778 \\
                &  RIDGE         & 0.784 & 0.697 & 0.639 & 0.783 & 0.789 & 0.772 & 0.834 & 0.802 & 0.789  & 0.784 & 0.782 & 0.789  \\
                &  RF            & 0.716 & 0.583 & 0.487 & 0.719 & 0.605 & 0.700 & 0.717 & 0.720 & 0.605  & 0.716 & 0.691 & 0.605 \\
                &  FMRS          & 0.780 & 0.676 & 0.568 & 0.780 & 0.706 & 0.769 & 0.727 & 0.797 & 0.706  & 0.780 & 0.780 & 0.706    \\
   \midrule
      Variable      &  \textbf{CSMR} & \textbf{0.999} & \textbf{0.950} & \textbf{0.538} & \textbf{1} & \textbf{0.980} & \textbf{1} & \textbf{0.956} & \textbf{1} & \textbf{0.980}  & \textbf{0.999} & \textbf{0.998} & \textbf{0.980}  \\
   Selection    &  ICC           & 0.500 & 0.332 & 0.339 & 0.500 & 0.461 & 0.500 & 0.500 & 0.500 & 0.461  & 0.500 & 0.496 & 0.461  \\
      (TPR)     &  FMRS          & 0.679 & 0.552 & 0.487 & 0.681 & 0.579 & 0.674 & 0.672 & 0.706 & 0.579  & 0.679 & 0.635 & 0.500   \\
   \midrule
    Variable    &  \textbf{CSMR} & \textbf{0.993} & \textbf{0.976} & \textbf{0.785} & \textbf{0.994} & \textbf{0.992} & \textbf{0.968} & \textbf{0.966} & \textbf{0.990} & \textbf{0.992}  & \textbf{0.993} & \textbf{0.992} & \textbf{0.992} \\
   Selection    &  ICC           & 0.972 & 0.957 & 0.669 & 0.973 & 0.870 & 0.735 & 0.966 & 0.972 & 0.870 & 0.972 & 0.953 & 0.870\\
    (TNR)       &  FMRS          & 0.499 & 0.680 & 0.758 & 0.504 & 0.512 & 0.500 & 0.502 & 0.506 & 0.512 & 0.499 & 0.515 & 0.500\\
   \midrule
   Sample       &  \textbf{CSMR} & \textbf{0.893} & 0.833 & \textbf{0.624} & \textbf{0.943} & \textbf{0.917} & \textbf{0.787} & 0.852 & \textbf{0.886} & \textbf{0.917}  & \textbf{0.893} & \textbf{0.908} & \textbf{0.917}   \\
    Clustering  &  ICC           & 0.887 & \textbf{0.838} & 0.549 & 0.941 & 0.879 & 0.787 & \textbf{0.878} & 0.881 & 0.879  & 0.887 & 0.903 & 0.879  \\
     (RI)       &  FMRS          & 0.501 & 0.546 & 0.624 & 0.502 & 0.513 & 0.502 & 0.501 & 0.501 & 0.513  & 0.501 & 0.502 & 0.513   \\
    \bottomrule
  \end{tabular}
  \label{tab:table}
  \end{minipage}}
\end{table*}

\section{Application to CCLE data}
\subsection{Description of the dataset}
Over the past three decades, the use of molecular data to inform drug discovery and development pipeline has generated huge excitement. Predicting the drug sensitivity becomes an integral part of the precision health initiative. Although earlier efforts successfully identified many new drug targets, the overall clinical efficacy of the developed drugs has remained unimpressive, owing in large part to the population heterogeneity, that is, different patients may have different disease causing factors, and hence drug targets. Here, we apply CSMR to study the patient heterogeneity in their response to different drug treatments, and select the most key molecular features that underlie the heterogeneous disease causes.

We collected gene expression data of 470 cell lines on 7902 genes, as well as the cell lines' sensitivity score to all 24 drugs, from the Cancer Cell Line Encyclopedia (CCLE) dataset \cite{barretina2012cancer}. The sensitivity score, or called the AUCC score, is defined as the area above the fitted dose response curve, and it has been shown to have better predictive accuracy of sensitivity to targeted therapeutic strategies than other measures, such as IC50 or EC50 \cite{jang2014systematic}. We applied all five methods on the dataset, where the drug sensitivity score was treated as response variable and the gene expressions as independent variables. Here, FMRS is not applicable as the feature dimension is too high while the sample size is too small, hence it is omitted from further analysis.  Our goal is to study the biological mechanism of possible heterogeneity in drug sensitivity, under the hypothesis that cells exhibit subgroup characteristics by selecting different genes that confer their different levels of drug sensitivity.

\begin{figure*} 
    \centering
    \includegraphics[width=16cm]{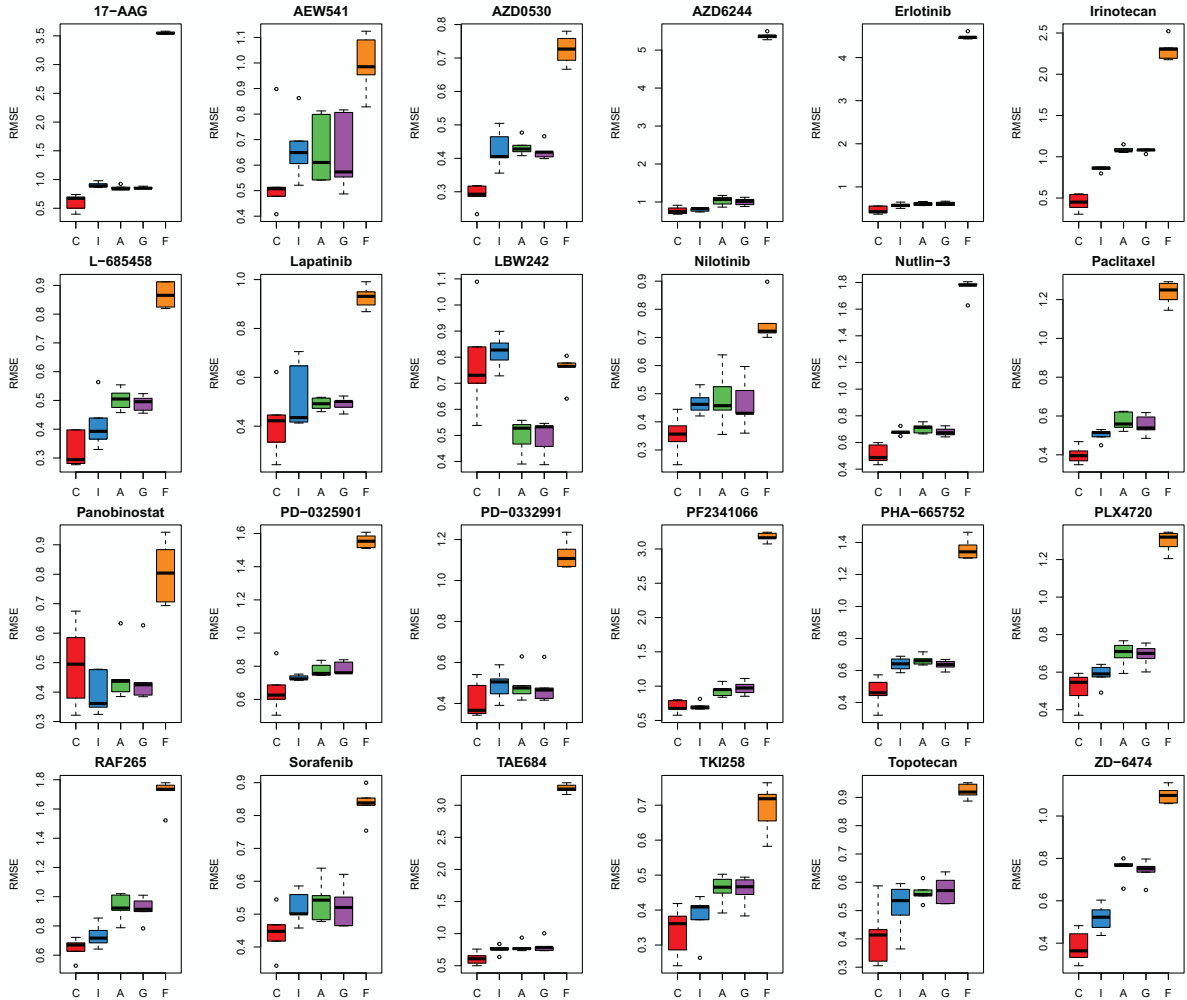}
    \caption{The distributions of the RMSE over 100 repetitions for the five methods, for the 24 drugs. The lower RMSE value, the better performance. `C',`I',`A',`G',`F' stand for `CSMR',`ICC',`LASSO',`RIDGE',`Random Forest'}
\end{figure*}

\begin{figure*} 
    \centering
    \includegraphics[width=16cm]{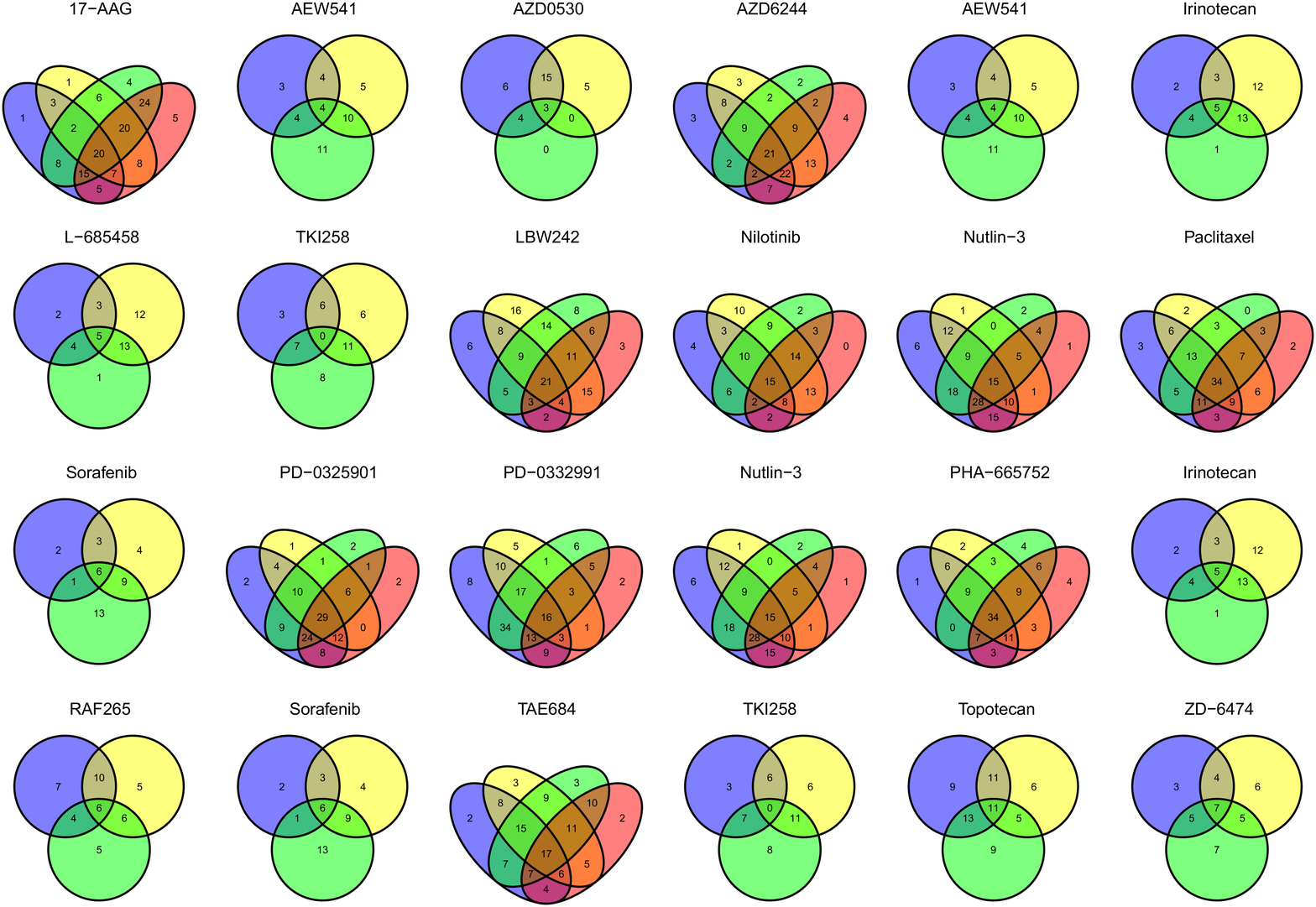}
    \caption{For each drug, the Venn diagram of the selected genes for different mixing components are shown. The numbers show the size of overlap between the gene sets.}
\end{figure*}

\subsection{Results}

We compare the performances of the five methods using cross validation. Basically, for each drug, we conduct a 5-fold cross validation by holding 80\% of the data as training, and 20\% as testing data, for each of the 100 repetitions. At each repetition, the 20\% testing data is used to independently evaluate the performance of each method. At the training phase, we start by fixing the hyper parameters involved in all methods. The penalty parameters for LASSO and RIDGE were selected by cross validation within the training samples. For RF, the default parameters were used in the function 'randomForest' of the package with the same name. For ICC, we used the selected component number as in its original paper \cite{li2019drug}. For CSMR, to select the best $K$, we performed both cross validation and the traditional BIC criteria introduced in Methods,  over a grid of $K=1,2,3,4,5,6$. We adopted the results from cross validation, as there is a lack of rigorous theoretical foundation for the validity of the traditional BIC under this high dimensional setting, and the data driven selection of cross validation seems more reasonable. The selected $K$ for BIC and cross validation using CSMR and used $K$ for ICC is summarized in Supplementary Table S1. With the hyper parameters fixed, we then conduct parameter estimations for each of the five methods using the training samples, and concludes the training phase. 

At the testing phase, the predicted and true drug sensitivity scores were examined in terms of their correlation, and residual mean squared error (RMSE). Note that this part of the testing data has never been used in the hyper parameter tuning or parameter estimation before. The distributions of RMSE and correlations over 100 repetitions for all the 24 drugs for all the five methods were shown in Figure 3 and Supplementary Figure S1, respectively. For 22 drugs, CSMR had the significantly smaller average RMSE, and was very close to the smallest RMSE for the rest of the two drugs; and we could make the same conclusions based on the correlation results as well. This demonstrated the consistent and robust performance of CSMR over the others. 

Among the five methods, RF had the poorest performance on the testing data, probably caused by model overfitting. LASSO and RIDGE worked much better than RF, probably due to its power in model selection. However, they performed significantly worse than ICC and CSMR in majority of the cases, which indicates the existence of population  heterogeneity and necessity of using mixture modeling. The performance of ICC is much worse than CSMR in most of the cases, which we believe is caused by the under-estimation of the population heterogeneity by ICC. In other words, the selection of $K$ in ICC is too conservative. In fact, according to cross validation, the number of distinct clusters given by CSMR for the drugs is either 3 or 4, while for ICC, the number of distinct clusters are determined to be less than 3 for half of the drugs. We believe that cross validation is a data driven approach for selection of $K$, and should be more reasonable than theoretically derived criteria. In the case of CCLE data, the samples are different types of  cells from very different experimental and genetic backgrounds, and it is expected that they would pick up different molecular mechanisms to deal with the attacks of the drugs. Hence, the cluster number given by CSMR is more realistic than ICC. It is wrothy of note that for those drugs that CSMR and ICC gave the same number of distinct clusters, namely Irinotecan, L-685458, Lapatinib, Paclitaxel, PD-0332991, PHA-665752 and TKI258, CSMR exhibited much smaller RMSE than ICC. 

Figure 4 demonstrated the Venn diagram of the selected genes for different components for each drug, and all the selected genes could be found in Supplementary Table S2. It could be seen that for the same drug, different clusters of cells indeed acquire different coping mechanisms, as seen by the different set of genes selected. This again confirms the high heterogeneous populations within the CCLE cohort. For each drug, we pooled all the selected genes together and conducted pathway enrichment analysis against 1,328 pathways collected in \cite{liberzon2011molecular}, and the top enriched pathways are shown in Supplementary Figure S2. Again, it could be seen that different responses to different drugs have been employed.

\section{Conclusions}

With the recent rapid evolution in genomic technologies, we have now entered a new phase, one in which it is possible to comprehensively characterize the molecular profiles of large population of subjects. Importantly, the development of sequencing technologies has been paired with a transition towards integrating molecular data with phenotypic data, such as in the electronic medical records. Such a synergy has the potential to ultimately facilitate the generation of a data commons useful for identifying relationships between molecular variations and their clinical presentations. Unfortunately, existing big data analysis tools for mining the information rich data commons has not been very impressive with regards to the overall transnational or clinical efficacy, owing in large part to the heterogeneous causes of disease. It is hence imperative to unveil the relationship between the molecular manifestations and the clinical presentations, while taking into account the possible heterogeneity of the study subjects.

In this paper, we proposed a novel supervised clustering algorithm using penalized mixture regression model, called CSMR, to deal with the challenges in studying the heterogeneous relationships between high dimensional molecular features to a phenotype. CSMR is capable of simultaneous stratification of the sample population and sparse feature-wise characterization of the subgroups. The algorithm was adapted from the classification expectation maximization algorithm, which offers a novel supervised solution to the clustering problem, with substantial improvement on both the computational efficiency and biological interpretability. Experimental evaluation on simulated benchmark datasets with different settings demonstrated that the CSMR can accurately identify the subspaces on which subset of features are explanatory to the response variables and the feature characteristics of the subspaces, and it outperformed the baseline methods. Application of CSMR on the heterogeneous CCLE dataset demonstrated the superior performance of CSMR over the others. On the CCLE dataset, CSMR is powerful in recapitulating the distinct subgroups hidden in the pool of cell lines with regards to their coping mechanisms to different drugs. CSMR also demonstrated the uniqueness of different subgroups for the same drug, as seen by the distinctly selected genes for the subgroups.

In summary, CSMR represents a big data analysis tool with the potential to bridge the gap between advancements in biotechnology and our understanding of the disease, and resolve the complexity of translating the clinical manifestations of the disease to the real causes underpinning it. We believe that such a tool will bring new understanding to the molecular basis of a disease, and could be of special relevance in the growing field of personalized medicine.

%
%

\bibliographystyle{unsrt}  
\bibliography{ref_bi}

\end{document}